# Investigating the Transition Region in Scanned Probe Images of the Cyclotron Orbit in Graphene

Sagar Bhandari,[1] Andrew Lin,[2] Robert Westervelt[1,2,*]

[1] School of Engineering and Applied Sciences, Harvard University, Cambridge, MA 02138, United States of America

[2] Department of Physics, Harvard University, Cambridge, MA 02138, United States of America

A cooled scanning probe microscope (SPM) has been used to image cyclotron orbits of electrons through high-mobility graphene in a magnetic field.[1-5] In a hBN-graphene-hBN device patterned into a hall bar geometry, the magnetic field focuses a current $I_i$ injected from one narrow contact into another narrow contact located an integer number of cyclotron diameters away, creating a voltage $V_c$. The degree of focusing is measured by the transresistance $R_m = V_c/I_i$. In SPM, the tip can either enhance or decrease conductance in the sample by deflecting electrons into or away from the second contact, respectively.[3,4] Our SPM images of magnetic focusing feature a region in which the tip transitions from enhancing to decreasing the conductance in the sample where the change in transresistance caused by the tip is equal to zero. In this paper, we investigate how the location of this region in the graphene sample changes as we modulate the electron density n and magnetic field B. By plotting line-cuts of the change in trans-resistance for different electron densities and magnetic fields, we identify trends in the inflection point where the tip changes from enhancing to decreasing the conductance in the sample. From the location of each transition region, we show that the cyclotron diameter of the electron trajectories can be obtained, and explain the trends in inflection point location for different electron densities and magnetic fields.

## 1. INTRODUCTION

The unique electronic and physical properties of graphene offer the potential for future electronic devices that take advantage of the ballistic transport of electrons in graphene. The fabrication of such devices is heavily contingent upon a fundamental understanding of the physics inherent to electron motion in graphene.

Electron motion in GaAs/AlGaAs heterostructures as well as in graphene and other two-dimensional materials has been imaged via scanning-probe microscopy (SPM).[1-5] In this technique, the charged tip of a scanning probe microscope is held just above the sample surface, creating an image charge inside the device that scatters electrons. By measuring the change in conductance while the tip is raster scanned above the sample, an image of electron motion can be obtained.[1-5]

SPM has been used to image electron flow through graphene in the magnetic focusing regime.[3] High-mobility graphene encapsulated between layers of hBN and etched into a hall-bar geometry features magnetic focusing of a current $I_i$ emitted from one contact into an adjacent contact an integer number of cyclotron orbits away. This conductance generates a voltage $V_c$, on the adjacent contact, and we measure the degree of magnetic focusing *via* the transresistance $R_m = V_c/I_i$.

Images of electron motion in the magnetic focusing regime are generated by plotting the transresistance $R_m$ vs. the tip position. In SPM technique, the tip decreases conductance in the sample by deflecting electrons away from the second contact, thus lowering $R_m$. Regions showing a strong decrease in $R_m$ indicate the paths of electrons, and we therefore image cyclotron orbits by displaying the drop in transresistance caused by the tip for each location in the sample.[3,4]

However, at a field lower than the focusing field, the tip increases transresistance close to the edge of the sample by deflecting electrons that would have scattered into the edge into the second contact.[3,4] In the magnetic focusing regime, our SPM images demonstrate the existence of an inflection point where the change in trans-resistance caused by the tip equals zero. This region marks where the tip changes from enhancing to reducing conductance in the sample.

In this paper, we present an analysis of how the location of this transition region changes with varying electron density n and magnetic field B. From the location of these transition regions, we show that the cyclotron diameter of the electron trajectories can be obtained and explain the observed trends in the spatial location of the transition region with varying n and B.

*Corresponding author Robert M Westervelt
Tel: +1 617 495 3296     Fax: +1 617 495 9837
email: westervelt@seas.harvard.edu

## 2. EXPERIMENTAL DETAILS

Cyclotron orbits in the magnetic focusing regime in a hBN-graphene-hBN structure were imaged (Fig. 1(a) using a home-built cooled scanning probe microscope at 4.2K.[3] A magnetic field B was applied transverse to the sample and modulated alongside the electron density n, which was tuned by varying the voltage between the doped-Si substrate which served as a back-gate and the graphene.

In our analysis, we looked at images in which the magnetic field was varied from 80 mT to 120 mT, while the electron density n was varied from $0.81 \times 10^{12}$ cm$^{-2}$ to $1.45 \times 10^{12}$ cm$^{-2}$ (Fig. 2(a) and 2(b)). From each image, we took a line-cut of the change in trans-resistance at Y = 1.25 μm and plotted the resulting curve (Fig. 2(c) and 2(d)).

We calculated the distance of each inflection point from the sample edge by finding the first zero point of each trans-resistance line-cut curve, which corresponds to the change-over point where the tip begins to deflect electrons away from rather than towards the contact. In addition, we plotted the location of each line-cut and inflection point on the images of magnetic focusing that were analyzed (Figs. 1 and 2).

## 3. RESULTS AND DISCUSSION

Magnetic focusing data from our group with overlaid line-cuts and inflection points are shown in Figs. 1 and 2. The blue region close to the sample edge in Fig. 1(a) corresponds to the region of positive change in transresistance in Fig. 1(b), and this region is where the tip is enhancing $R_m$ by deflecting edge-bound electrons into the sample edge. Likewise, the red and yellow region in the center of the sample in Fig. 1(a) corresponds to the region of drop in $R_m$ in Fig. 1(b), where the tip is deflecting electrons away from the second contact. The black semicircular region bounding the blue and red regions in Fig. 1(a) marks where the change in transresistance caused by the tip is near zero.

Plots of the change in transresistance ΔR and the location of the inflection point where the change in transresistance is zero are shown in Fig. 2(c) and 2(d). A negative correlation is seen between the inflection point distance from the sample edge and the electron density for constant magnetic field B = 100 mT, while a positive correlation is clearly shown between the inflection point distance from the sample edge and the magnetic field for constant electron density n = $0.97 \times 10^{12}$ cm$^{-2}$ (Fig. 2(c)-(d)).

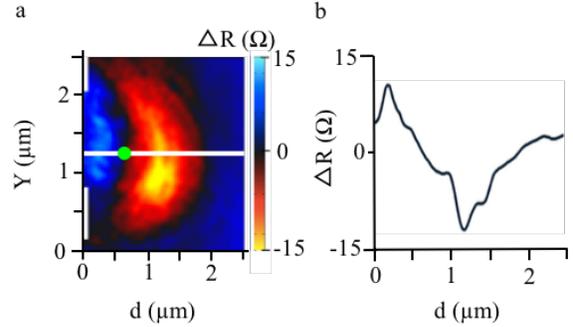

**Fig. 1:** (a) Image of magnetic focusing with plotted line-cut and inflection point at magnetic field = 100 mT and n = $0.97 \times 10^{12}$ cm$^{-2}$. The white line indicates the location of the line-cut at Y = 1.25 μm, while the green dot indicates the location of the inflection point where the trans-resistance changes from positive to negative. (b) Line-cut plot of transresistance vs. distance from the sample edge for magnetic focusing image in Fig. 1(a), where d is the distance from the edge of the sample.

The diameters of the semicircular transition region for each magnetic focusing image are plotted in Fig. 3. Figure 3(a) displays an example image with the transition region circle highlighted in green; the diameter of each circle as indicated by the blue arrow was measured for the magnetic focusing images.

The diameter of each semicircular region was compared to the theoretical cyclotron diameter for each set of experimental conditions. In graphene, the cyclotron diameter $d_c = 2m^*v_F/eB$ increases with electron density as $n^{1/2}$ by virtue of the contribution of the dynamical mass of graphene, $m^* = \hbar(\pi n)^{1/2}/v_F$. Plots of the calculated cyclotron diameter and the diameter of the semicircular transition region are shown in Figs. 3(b) and 3(c).

Figures 3(b) and 3(c) demonstrate a clear positive correlation between the diameter D of the semicircular transition region and electron density n and a clear negative correlation between D and the magnetic field B. Additionally, the diameter D of the semicircular transition region is correlated closely to the predicted cyclotron diameter for varying magnetic fields and electron density (Fig. 3(b) and (c)). These results thus show that the cyclotron diameter can be directly obtained from the diameter of the semicircular transition region for magnetic focusing imaged using our SPM technique.

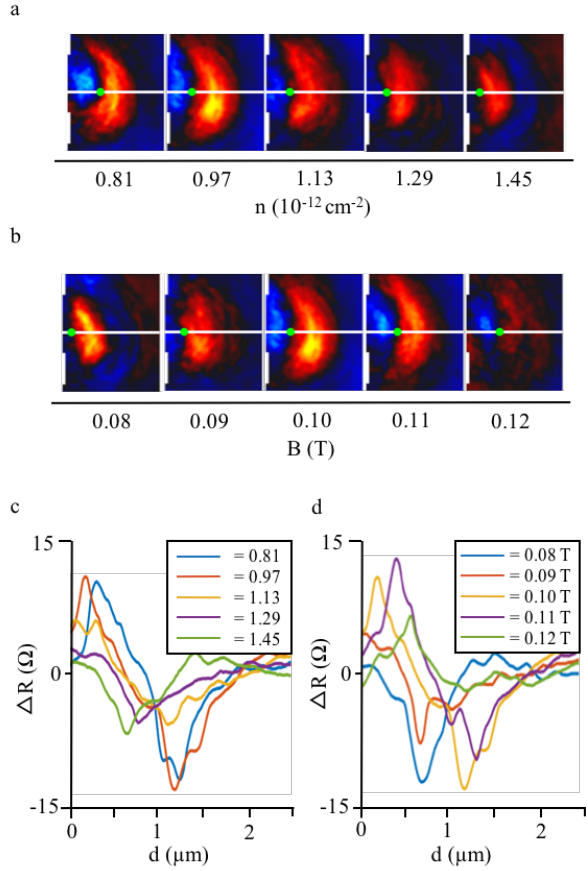

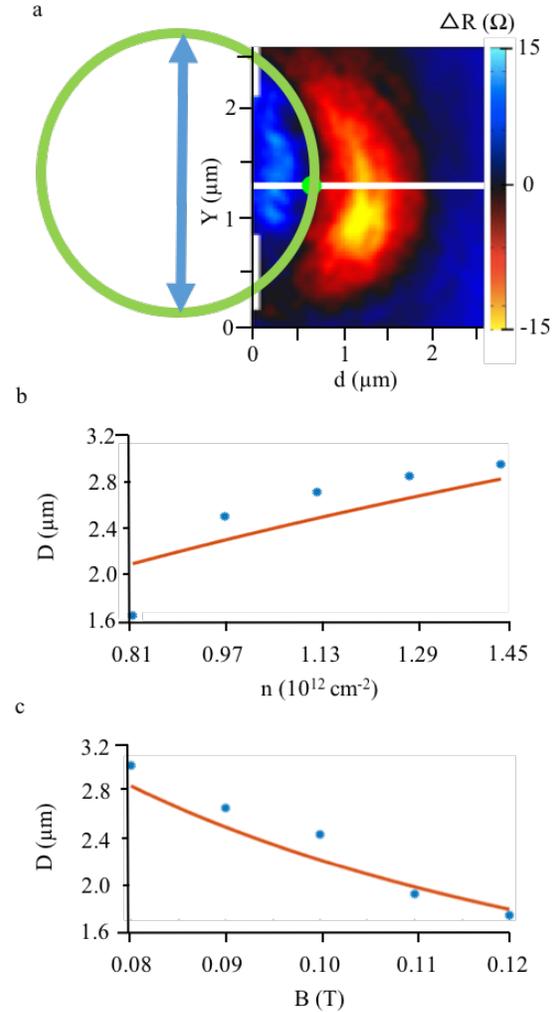

**Fig. 2**. Plots of line-cuts and transresistance $R_m$ for varying electron density n and magnetic field B. (a) Magnetic focusing images for varying electron densities n (in units of $10^{12}$ cm$^{-2}$) with magnetic field B = 100 mT. (b) Magnetic focusing images for varying magnetic field with n = $0.97 \times 10^{12}$ cm$^{-2}$. (c) Line graph of transresistance change $\Delta R$ vs. the distance d of the inflection point from the edge of the sample for varying electron densities n in units of $10^{12}$ cm$^{-2}$ at B = 100 mT. (d) Line graph of $\Delta R$ for varying magnetic fields B at electron density n = $0.97 \times 10^{12}$ cm$^{-2}$.

**Fig. 3.** Measured diameters D of the semicircular transition region plotted *vs.* the calculated cyclotron radius for varying electron density n and magnetic field B. (a) Image of magnetic focusing at B = 100 mT and n = $0.97 \times 10^{12}$ cm$^{-2}$ with the semicircular transition region highlighted as a green circle. The diameter D of the transition region is highlighted as the blue arrow. (b) Plot of the transition region diameter D *vs.* electron densities. The theoretical cyclotron diameter is plotted as the red line. (c) Plot of D *vs.* B. The theoretical cyclotron diameter is plotted as the red line.

The relationship between the cyclotron diameter $d_c$ and the diameter of the semicircular transition region explains how the inflection point distance changes for varying electron density and magnetic fields. As B increases, the cyclotron diameter decreases $d_c = 2\hbar(\pi n)^{1/2}/eB$, a result closely borne out by the experimental data in Fig. 3(c), which shows a decrease in the diameter D of the transition region. For higher magnetic fields, the tip enhances transmission further from the edge of the sample due to the cyclotron diameter of electrons being less than the contact spacing.

Likewise, as the density n increases, the cyclotron diameter of electrons in the sample increases, which is also reflected in the experimental data in Fig. 3(c). This similarly corresponds to an increase in the diameter D of the transition region. For higher electron densities, the tip decreases transmission closer to the sample edge due to the cyclotron diameter of electrons being greater than the contact spacing.

## 4. CONCLUSION

The results presented in this paper clearly demonstrate that the tip can deflect the flow of electrons either into or away from the contact and suggest the design of future ballistic devices that guide electron paths in graphene, aided by visualization with our cooled SPM. It has been demonstrated that information such as an accurate measurement of the diameter of cyclotron orbits can be extracted directly from the SPM images. The properties of graphene offer much potential in the development of future devices for probing the boundaries of physics in two-dimensional materials as well as in other areas of science.

**Acknowledgments:** Supported by Dept. of Energy grant DE-FG02-07ER46422 and Air Force Office of Scientific Research contract FA9550-13-1-0211.